\documentclass[12pt]{iopart}

\usepackage{epsfig}

\begin{document}

\title{Superconductivity and abnormal pressure effect in Sr$_{0.5}$La$_{0.5}$FBiSe$_{2}$ superconductor}

\author{Lin Li$^{1}$, Yongliang Xiang$^{1}$, Yihong Chen$^{1}$, Wenhe Jiao$^{2}$, Chuhang Zhang$^{2}$, Li Zhang$^{3}$, Jianhui Dai$^{1}$, Yuke Li$^{1}$\footnote[1]{Electronic address: yklee@hznu.edu.cn}}

\address{$^{1}$Department of Physics and Hangzhou Key Laboratory of Quantum Matter, Hangzhou Normal University, Hangzhou 310036, China}
\address{$^2$Department of Physics, Zhejiang University of Science and technology  Hangzhou 310023, China}
\address{$^3$Department of Physics, China Jiliang University, Hangzhou 310018, China}

\date{\today}

\begin{abstract}

Through the solid state reaction method, we synthesized a new
BiSe$_2$-based superconductor Sr$_{0.5}$La$_{0.5}$FBiSe$_{2}$ with
superconducting transition temperature \emph{T$_{c}$}$\approx$3.8 K.
A strong diamagnetic signal below $T_c$ in susceptibility $\chi(T)$
is observed indicating the bulk nature of superconductivity.
Different to most BiS$_2$-based compounds where superconductivity
develops from a semiconducting-like normal state, the present
compound exhibits a metallic behavior down to $T_c$. Under weak
magnetic field or pressure, however, a remarkable crossover from
metallic to insulating behaviors takes place around $T_{min}$ where
the resistivity picks up a local minimum. With increasing pressure,
\emph{T$_{c}$} decreases monotonously and $T_{min}$ shifts to high
temperatures, while the absolute value of the normal state
resistivity at low temperatures first decreases and then increases
with pressure up to 2.5 GPa. These results imply that the electronic
structure of Sr$_{0.5}$La$_{0.5}$FBiSe$_{2}$ may be different to
those in the other BiS$_{2}$-based systems.

\end{abstract}
\pacs{74.70.Dd,74.25.F-; 74.62.Fj; 74.25.Dw}
\maketitle

\section{\label{sec:level1}Introduction}
Exotic superconductivity has been discovered in materials with
layered crystal structures, such as the high-\emph{T$_{c}$}
cuprates\cite{CuO} and the Fe-based superconductors\cite{FeO} where
the high-$T_{c}$ or unconventional superconductivity is believed to
be caused by the reduced dimensionality and electronic correlations.
Recently, a new layered compound Bi$_{4}$O$_{4}$S$_{3}$ with
\emph{$T_{c}$} $\approx$ 8.6 K\cite{BOS,BOS2} has triggered
intensive research interests, leading to a class of  BiS$_2$-based
superconductors including
\emph{Ln}O$_{1-x}$F$_{x}$BiS$_{2}$(\emph{Ln}=La, Ce, Pr, Nd) with
\emph{T$_{c}$} up to $~$10 K\cite{LaFS,NdFS,NdFS2,LaFS2,CeFS,PrFS}.
Similar to the CuO$_{2}$ plane in cuprates and the
Fe$_{2}$\emph{An}$_{2}$ (\emph{An} = P, As, Se) layers in pnictides,
the common BiS$_{2}$ layer in the BiS$_{2}$-based compounds is
expected to play a key role in search for new superconductors by
intercalating various block layers, for example, the
Bi$_{4}$O$_{4}$(SO$_{4}$)$_{1-x}$ or [Ln$_{2}$O$_{2}$]$^{2-}$
layers. Following this idea, through the replacement of the LaO
layer by the SrF block, a new class of BiS$_{2}$-based
superconductors Sr$_{1-x}$\emph{Ln}$_{x}$FBiS$_{2}$(\emph{Ln}=La,
Ce) with $T_{c}$ $\approx $ 2.8 K has been
studied\cite{LiSrF,LiSrCe,SSTLi}. By first principle calculations
the parent compounds  SrF(LnO)BiS$_2$ are found to be band
insulators without detectable antiferrimagnetic transition or
structure phase transition\cite{es}. Superconductivity can be
induced by electron doping into lattice\cite{LaFS} and enhanced by
pressure\cite{SCT}.

To obtain higher $T_{c}$ in these materials, various chemical
substitutions have been attempted to alter the structural
instability\cite{LaFS,LiSrF,LaTixBiS2}. Among these, the isovalent
substitution is a clean method to supply chemical pressure. Very
recently, an isostructural LaO$_{1-x}$F$_{x}$BiSe$_{2}$ compound has
been reported to exhibit enhanced-superconductivity with $T_{c}$ of
3.5 K\cite{LaOSe,LaOSe2} compared to the low-$T_{c}$ phase in
LaO$_{1-x}$F$_{x}$BiS$_{2}$\cite{LaFS,SCT}. Although the ARPES
experiments\cite{ARPSe} have suggested that the electronic structure
and Fermi surface are quite similar in both compounds, the normal
state of LaO$_{1-x}$F$_{x}$BiSe$_{2}$ shows the metallic
behavior\cite{LaOSe} in contrast to the semiconducting behavior of
LaO$_{1-x}$F$_{x}$BiS$_{2}$. On the other hand, the applied physical
pressures in BiS$_2$-based compounds always induce a structure phase
transition from tetragonal to monoclinic \cite{XRD}, leading to an
abrupt improvement from low-$T_{c}$ to high-$T_{c}$\cite{SCT}.  For
LaO$_{1-x}$F$_x$BiSe$_2$, the reported pressure effect is to
suppress the low-$T_{c}$ but enhance the high-$T_{c}$\cite{WenHH}.
Up to now, most studies including electronic structure\cite{es},
superconducting transition temperature\cite{SCT} and the pairing
symmetry\cite{HJP,Yildirim,Martins,usR,ARPES,ARPES2} have been
mainly focused on the BiS$_{2}$-based system, but superconductivity
seems to be still under debate.

In the Letter, we report the successful synthesis of a novel
La-doped Sr$_{0.5}$La$_{0.5}$FBiSe$_{2}$ sample. The compound is
iso-structural to SrFBiS$_{2}$ with the P4/nmm space group as
confirmed by the XRD pattern measurement. Both the sharp
superconducting transition in $\rho(T)$ and strong diamagnetic
signals in $\chi(T)$ confirm the bulk superconductivity. In contrast
to most of the BiS$_2$-based compounds where the normal state
exhibits a semiconducting behavior, Sr$_{0.5}$La$_{0.5}$FBiSe$_{2}$
exhibits a metallic behavior down to $T_{c}$. Interestingly, even a
weak magnetic field or a weak pressure can induce a crossover from
metallic to insulating behaviors in the normal state. In particular,
by increasing pressure, superconductivity is quickly suppressed.
Accompanied with the decrease of $T_{c}$, the normal state resistivity first
decreases and then increases with pressure. In any cases, the
resistivity has a local minimum at $T_{min}$ where the crossover
from metallic to insulating behaviors takes place. We find that
$T_{min}$ gradually shifts to high temperature with increasing
pressure or field. All these observations imply that the
superconducting mechanism of the present system may be distinct from
that of the BiS$_{2}$-based superconductors.

\section{\label{sec:level1}Experimental}

The polycrystalline sample of Sr$_{0.5}$La$_{0.5}$FBiSe$_{2}$ used
in this study was synthesized by the two-step solid state reaction
method. La$_{2}$Se$_{3}$ was pre-synthesized by reacting
stoichiometric Se powders and La pieces at 1273 K for 15 hours. The
as-grown La$_{2}$Se$_{3}$ and the powders of SrSe, SrF$_{2}$, Bi,
and Se as starting materials were weighted according to their
stoichiometric ratio and then fully ground in an agate mortar. The
mixture of powders was then pressed into pellets, heated in an
evacuated quartz tube at 1073 K for 10 hours and finally
furnace-cooled to room temperature.

Crystal structure characterization was performed by powder X-ray
diffraction (XRD) at room temperature using a D/Max-rA
diffractometer with Cu K$_{\alpha}$ radiation and a graphite
monochromator. Lattice parameters were calculated from least-squares
fitting routine using Rietveld fitting. The (magneto)resistivity
under several magnetic fields was measured with a standard
four-terminal method covering temperature range from 2 to 300 K in a
commercial Quantum Design PPMS-9 system and Oxford He$^{3}$-16T
system. The temperature dependence of d.c. magnetization was
measured on a Quantum Design SQUID-VSM-7T. Measurement of
resistivity under pressure was performed up to 2.5GPa on PPMS-9T by
using HPC-33 Piston type pressure cell with the Quantum Design DC
resistivity and AC transport options. Hydrostatic pressures were
generated by a BeCu/NiCrAl clamped piston-cylinder cell. The sample
was immersed in a pressure transmitting medium (Daphne Oil) covered
with a Teflon cell. Annealed Au wires were affixed to contact
surfaces on each sample with silver epoxy in a standard four-wire
configuration.

\section{\label{sec:level1}Results and Discussion}

\subsection{\label{sec:level1}{Superconductivity}}

Fig. 1 shows the powder XRD patterns and the Rietveld structural
refinement of the Sr$_{0.5}$La$_{0.5}$FBiSe$_{2}$ sample. The main
diffraction peaks can be well indexed based on a ZrCuSiAs-type
crystal structure with the P4/nmm space group except for two
impurity phases, the Bi$_{2}$Se$_{3}$ and a starting material Bi.
The refined lattice parameters are extracted to be $a=$ 4.1697{\AA} and $c=$
13.9422{\AA}, which are larger than those of
Sr$_{0.5}$La$_{0.5}$FBiS$_{2}$ with $a=$ 4.0820 {\AA} and $c=$
13.8025 {\AA}, respectively\cite{LaFS}.

Temperature dependence of resistivity ($\rho$) for
Sr$_{0.5}$La$_{0.5}$FBiSe$_{2}$ under ambient pressure is plotted in
Fig. 2. The inset of figure 2 shows the magnetic susceptibility
under ZFC(Zero-Field cooling) and FC(Field cooling) modes with a
magnetic field of 5 Oe. The strong diamagnetic signals are observed
below 3.5 K. The estimated volume fraction of magnetic shielding
from ZFC data is over 90\%, bearing out the bulk superconductivity
in our sample. The resistivity displays a metallic behavior in the
whole temperature region above $T_{c}$, and, a linear
temperature-dependence above 100 K. Such feature is in contrast to
the semiconducting feature observed in
Sr$_{0.5}$La$_{0.5}$FBiS$_{2}$\cite{LiSrF} and
LaO$_{0.5}$F$_{0.5}$BiS$_{2}$\cite{LaFS}.
Noted that the single crystal NdO$_{0.5}$F$_{0.5}$BiS$_{2}$\cite{NdOFBiS} also exhibits the metallic behavior above $T_c$.
This result seems to be a tendency of the metallic normal state in the BiS$_2$- and BiSe$_2$-based superconductors.
Upon further cooling, a sharp superconducting transition with
\emph{T$_{c}$} of 3.8 K, which is sizably higher than that of
Sr$_{0.5}$La$_{0.5}$FBiS$_{2}$ with $T_{c}$ $\sim$ 2.8 K, can be
clearly seen. Considering the relatively larger radius of Se ion
than that of S ion, the result seems to imply that the negative
chemical pressure effect may enhance the superconductivity in the
Sr$_{0.5}$La$_{0.5}$FBiSe$_{2}$ system.

Fig. 3(a) shows the low-temperature magnetoresistivity under various
magnetic fields below 6 K for the Sr$_{0.5}$La$_{0.5}$FBiSe$_{2}$
sample. The sharp superconducting transition with fully vanished
resistivity at about 3.5 K is clearly seen at zero-magnetic field,
suggesting the good quality of the poly-crystalline sample. A
relatively weak $H$ ($\sim$0.8 T) suppresses $T_{c}$ drastically,
and induces an non-zero resistivity above 2 K, implying a low
Meissner field due to the pinning of flux. The inset shows the
temperature dependence of the upper critical field
$\mu_0$$H_{c2}(T)$, determined by using the 90\% normal state
resistivity criterion. The $H_{c2}$ at zero temperature estimated by
using the Werthamer-Helfand-Hohenberg(WHH) formula ${H_{c2}(T) =
-0.69T_c|{\frac{\partial{H_{c2}}}{\partial{T}}|_{T_c}}}$ is about
5.5 T for $T_c^{onset}$. This value is rather large compared to that
of the Sr$_{0.5}$La$_{0.5}$FBiS$_{2}$ system\cite{LiSrF}.

A logarithmic plot of the magnetoresistivity vs. temperature below
50 K with the applied magnetic field up to 9 T is shown in Fig.
3(b). Clearly, the resistivity shows a metallic behavior under zero
field. Small magnetic fields cause a slight upturn above $T_{c}$ and
broaden the superconducting transition. With increasing magnetic
field $T_{c}$ shifts to lower temperature, while the value of
resistivity gradually increases so that the upturn feature becomes
more prominent. At higher field up to 9 T, superconductivity almost
vanishes, instead, the $\rho(T)$ curves show a metal to
semiconductor crossover around $T_{min}$, followed by a near
\emph{logT}-dependent feature at lower temperatures.
The similar behavior was also observed in the
LaO$_{0.5}$F$_{0.5}$BiSe$_{2}$ crystal at 2.0 GPa\cite{WenHH}.
As a result, the phase diagram in terms of  magnetic field and
temperature is mapped in Fig. 4.

\subsection{\label{sec:level1}{Pressure effect}}

It is known that pressure is an effective method to tune the lattice
structures and the corresponding electronic states without
introducing more disorders. We performed the resistivity measurement
for several different pressures, shown in Fig. 5. The inset shows
the close view of resistivity transition below $T_{c}$ at various
pressures. A relatively weak pressure can sizably reduce $T_{c}$ but
only slightly broaden the superconducting transition. Further
increasing pressures, $T_{c}$  shifts to lower temperatures quickly
while the superconducting transition remains rather sharp. At higher
pressures up to 2.5 GPa, a slight drop due to superconducting
transition can be distinguished below 2.2 K. The fact that $T_c$
decreases monotonously with pressure in the present
Sr$_{0.5}$La$_{0.5}$FBiSe$_{2}$ sample is in contrast to other
BiS$_2$-based superconductors such LnO$_{1-x}$F$_x$BiS$_2$ and
Sr$_{1-x}$Ln$_x$FBiS$_2$ systems\cite{maple2,Awana2}, where the
$T_c$ is enhanced to 10 K by pressure. In the
LaO$_{1-x}$F$_x$BiSe$_2$ and Eu$_3$F$_4$Bi$_2$S$_4$
systems\cite{WenHH,LaOFBiSe,EuF3244}, while the superconducting
phase with low-$T_{c}$ is relatively unchanged, the one with a
high-$T_{c}$ is enhanced with $T_{c}\approx$ 10 K up to 2.5 GPa.

Fig. 6 displays a close view of temperature dependence of
resistivity at several representative pressures.
Starting from 0.9 GPa, a resistivity upturn above $T_c$  is induced
by pressure. The upturn feature becomes more pronounced with
increasing pressures, resulting in a clear crossover from metallic
to semiconducting behaviors around $T_{min}$ where the resistivity
takes a local minimum. Apparently, $T_{min}$ shifts toward higher
temperatures with pressures. It is noted that the value of
$\rho_{300 K}$ decreases with pressure, while $\rho_{10 K}$ first
decreases below 1.3 GPa and then increases above 1.5 GPa.
Consequently, while superconductivity is suppressed by pressure, the
$log(T)$-dependence of resistivity in the normal state emerges. The
region with this insulating feature increases quickly with pressure.
Therefore compared with the pressure effect on other mentioned
materials the pressure effect in the Sr$_{0.5}$La$_{0.5}$FBiSe$_{2}$
compound is abnormal.  The measurement under further higher
pressures should be highly desirable in the future in order to
clarify whether the superconductivity in Sr$_{0.5}$La$_{0.5}$FBiSe$_{2}$ could be completely killed by
pressure.

The phase diagram of pressure vs. temperature is summarized in Fig.
7. In the superconducting state, $T_c$ in the present system
decreases monotonously with pressure up to 2.5 GPa, in contrast to
the universal features in other BiS$_2$-based superconductors. In the
normal state, the sample shows a highly metallic character in the
whole temperature at ambient pressure, but undergoes a crossover
from metallic to insulating behaviors when the physical pressure is
beyond 0.9 GPa. With pressure up to 2.5 GPa, the superconductivity
is suppressed and the sample becomes more insulating at low
temperatures. Recall that $T_c$ is enhanced in most BiS$_2$- or
BiSe$_2$-based compounds as typically in LaO$_{1-x}$F$_{x}$BiSe$_2$
where $T_c$ is enhanced to 6.5 K at 2.37 GPa \cite{WenHH}. The
opposite pressure effect in our measured sample though with the
similar BiS$_2$ layered structure may suggest a rather different
electronic band structure.

\section{\label{sec:level1}Conclusion}
In summary, we synthesized the Sr$_{0.5}$La$_{0.5}$FBiSe$_{2}$
polycrystalline sample. The resistivity vanishes below 3.8 K, which
together with strong diamagnetic signals in magnetization data,
confirming the bulk superconductivity. In contrast to most of the
BiS$_2$-based compounds where superconductivity is developed from
the background of a semiconducting-like normal state, the normal
state of Sr$_{0.5}$La$_{0.5}$FBiSe$_{2}$ exhibits a metallic
behavior down to $T_{c}$.  Under magnetic field or pressure, a
crossover from metallic to semiconducting behaviors is induced, and
the superconductivity is suppressed accordingly. While the $T_{c}$
decreases with increasing monotonously, the absolute value of the
normal state resistivity first decreases and then increases with
pressure. All these features are in contrast to the previously known
BiS$_2$-based superconductors.

While the semiconducting behavior in the normal state of most
BiS$_2$ and BiSe$_2$ remains one of the puzzling issue in connection
with the unconventional superconductivity, the opposite situation in
the present compound, namely, the crossover from the metallic to
semiconducting behaviors in the normal in the presence of
 magnetic field or pressure is rather unusual, pointing to a
 possible different mechanism of superconductivity in this family of
 materials. Such feature reminds us of a related single crystal compound
 Nd(O,F)BiS$_2$\cite{WenHH2}, where the normal state exhibits the field-induced
 semiconducting behavior above $T_{c}$. This feature was attributed to
 the possible pseudo-gap phase extending to relatively higher temperatures as evidenced
 by the unusual superconducting fluctuations seen in the scanning tunneling
 spectroscopy (STS) experiment\cite{WenHH2}. We thus expect that the superconducting mechanism of the present
system may be also quite unique and deserve further investigations such
as by ARPES, NMR or STS experiments in future.

\section*{Acknowledgments}

Y. Li would like to thank Z. Xu and G. H. Cao for helpful discussions. This work is supported by the National Science
Foundation of China (Grant No. 11274084 and 61376094) and National Training Programs of Innovation and Entrepreneurship for Undergraduates (201510346011).

\pagebreak[4]

\begin{figure}
\includegraphics[width=14cm]{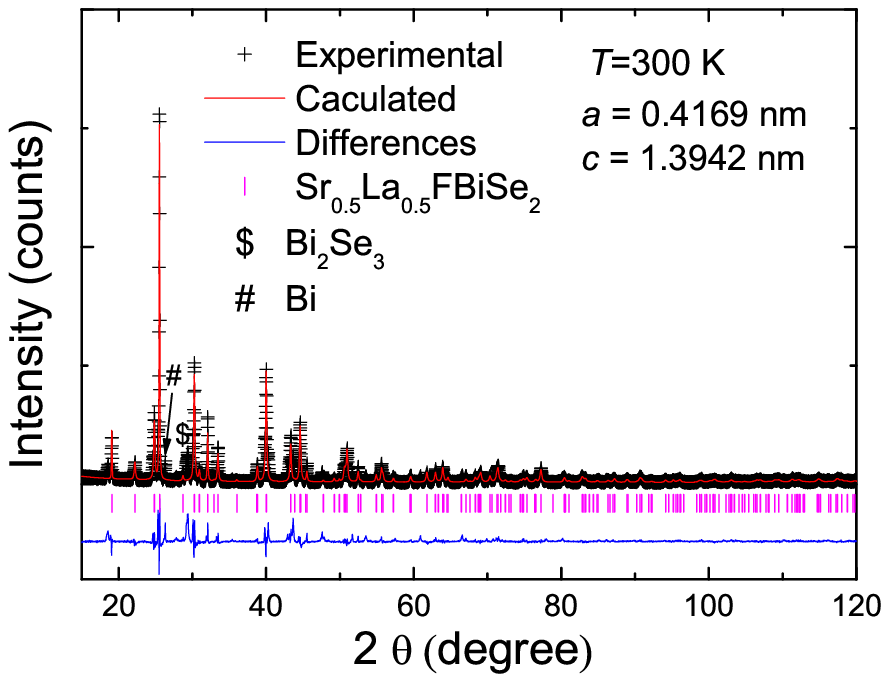}
\caption{\label{Fig.1} (color online). Powder X-ray diffraction
patterns and the Rietveld refinement profile for
Sr$_{0.5}$La$_{0.5}$FBiSe$_{2}$ sample at room temperature. The \$
and \# peak positions designate the impurity phases of
Bi$_{2}$S$_{3}$ and Bi, respectively. }
\end{figure}

\begin{figure}[h]
\includegraphics[width=14cm]{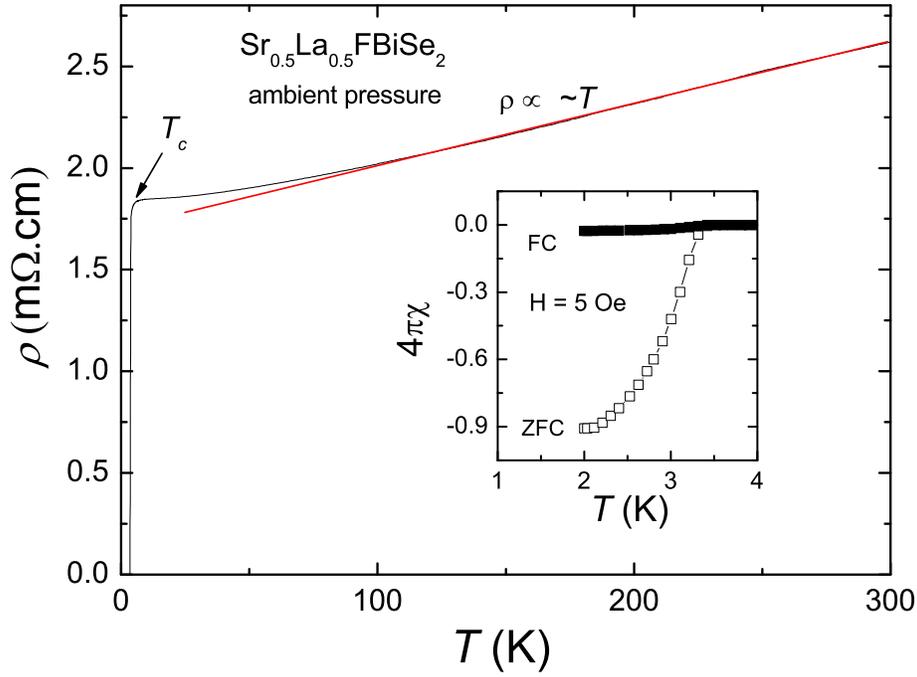}
\caption{\label{label} (a) Temperature dependence of resistivity for the polycrystalline Sr$_{0.5}$La$_{0.5}$FBiSe$_{2}$ sample under ambient pressure. The inset shows the magnetic susceptibility of Sr$_{0.5}$La$_{0.5}$FBiSe$_{2}$ under both ZFC and FC modes.}
\end{figure}

\begin{figure}[h]
\includegraphics[width=8cm]{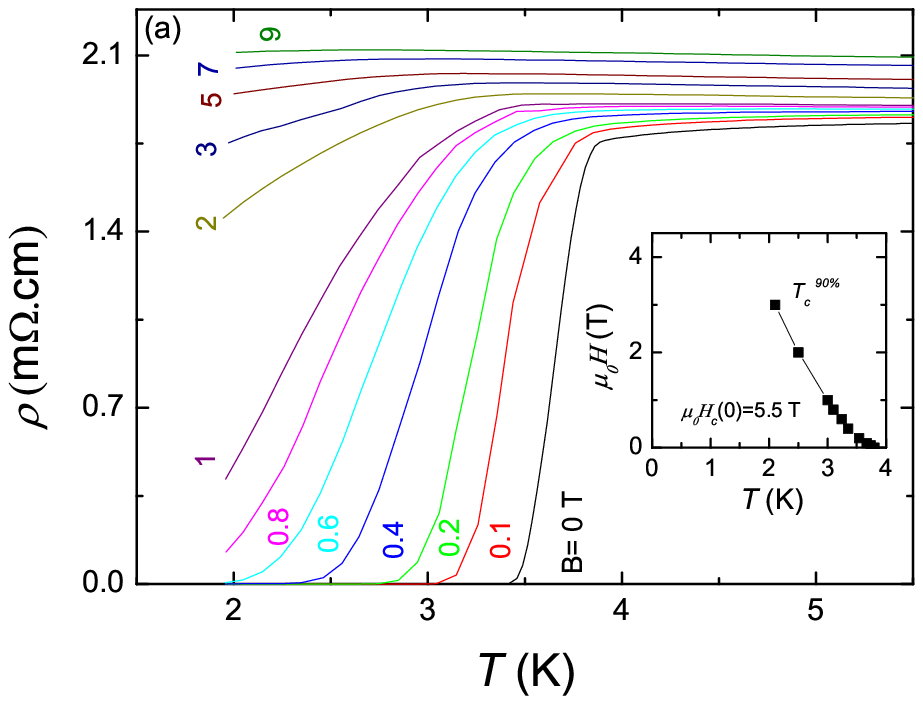}
\includegraphics[width=8cm]{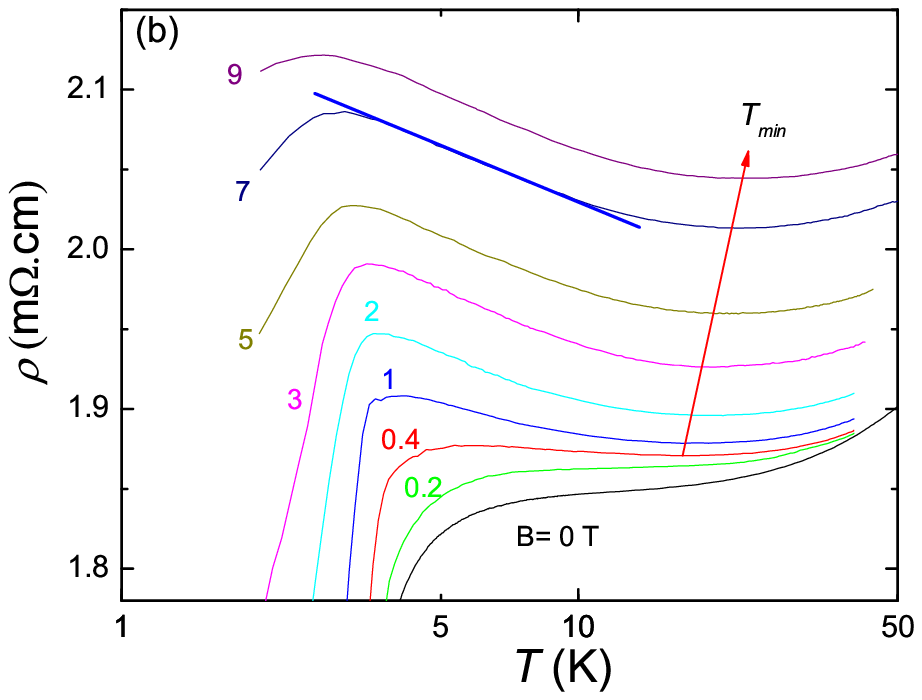}
\caption{\label{label}(a)Temperature dependence of resistivity
($\rho$) around $T_c$ under magnetic fields up to 9 T for the
Sr$_{0.5}$La$_{0.5}$FBiSe$_{2}$, the inset shows the $H_{c2}$ data.
(b) An enlarged plot of the temperature dependence of
magnetoresistivity.}
\end{figure}

\begin{figure}[h]
\includegraphics[width=14cm]{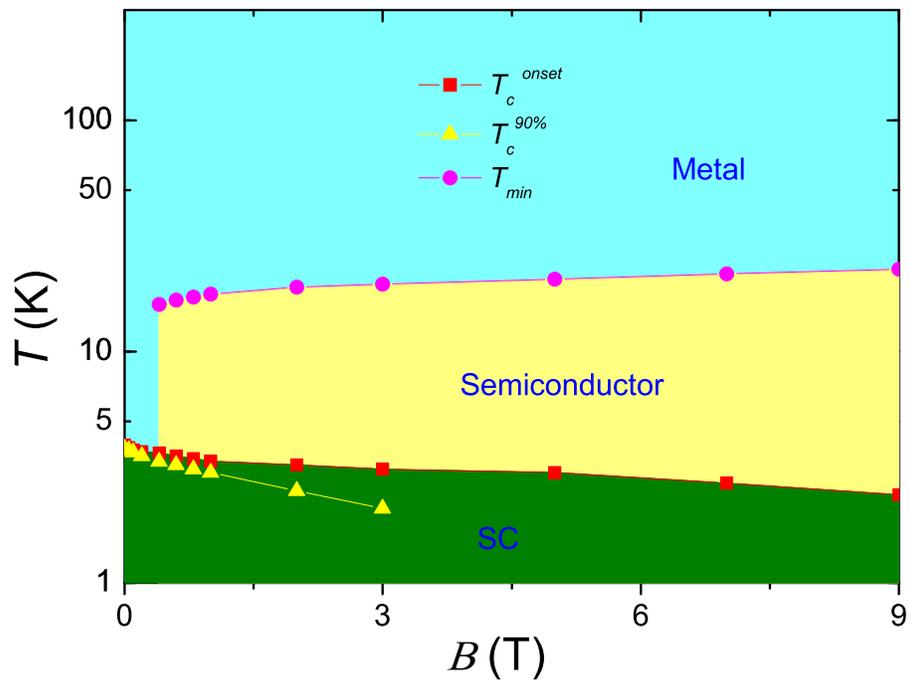}
\caption{\label{label} The phase diagram of 
temperature versus magnetic field for the Sr$_{0.5}$La$_{0.5}$FBiSe$_{2}$.}
\end{figure}

\begin{figure}[h]
\includegraphics[width=14cm]{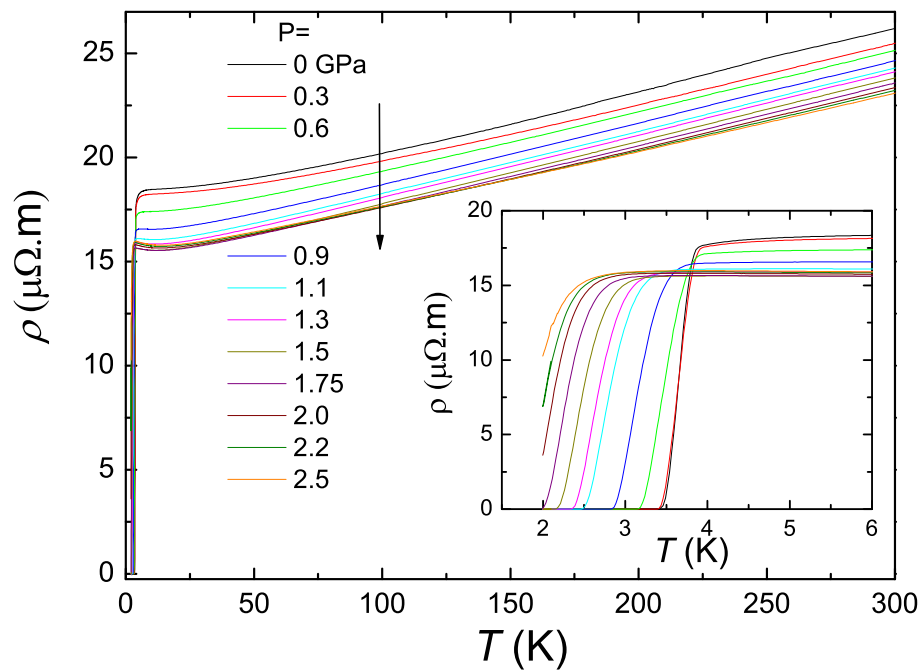}
\caption{\label{label} Temperature dependence of resistivity for
Sr$_{0.5}$La$_{0.5}$FBiSe$_{2}$ sample at various pressures. The
inset shows an enlarged view of resistivity below $T_c$. }
\end{figure}

\begin{figure}[h]
\includegraphics[width=14cm]{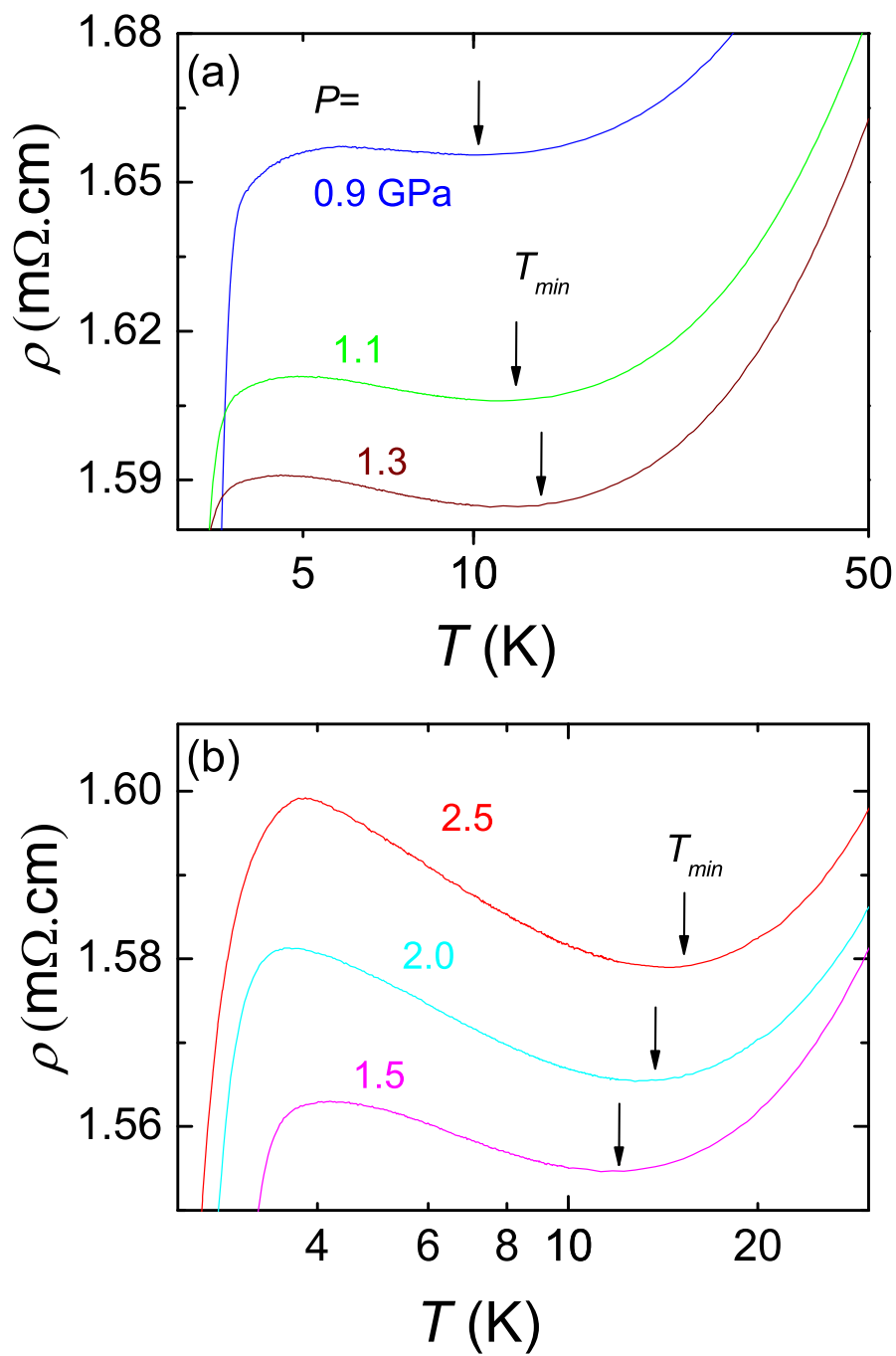}
\caption{\label{label} An enlarged view of temperature vs.
resistivity at several representative pressures for the
Sr$_{0.5}$La$_{0.5}$FBiSe$_{2}$ sample.}
\end{figure}

\begin{figure}[h]
\includegraphics[width=14cm]{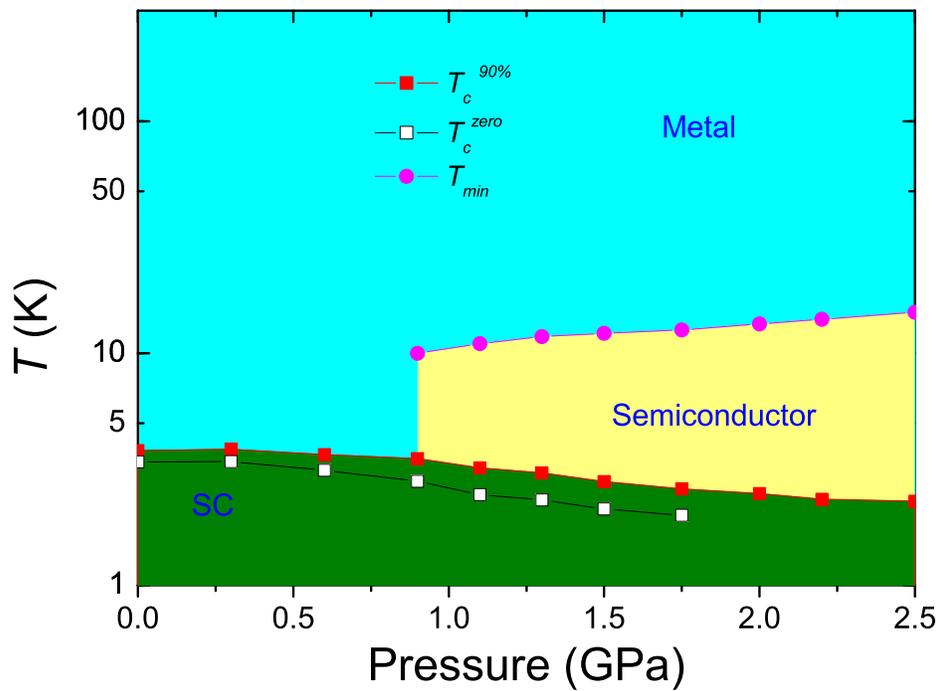}
\caption{\label{label} The phase diagram in terms of pressure and
temperature for the Sr$_{0.5}$La$_{0.5}$FBiSe$_{2}$ sample.}
\end{figure}

\end{document}